# Quasiparticle interference of $C_2$-symmetric surface states in LaOFeAs parent compound

Xiaodong Zhou[1,*], Cun Ye[1,*], Peng Cai[1], Xiangfeng Wang[2], Xianhui Chen[2], and Yayu Wang[1,†]

[1]Department of Physics, Tsinghua University, Haidian, Beijing 100084, P.R. China

[2]Hefei National Laboratory for Physical Science at Microscale and Department of Physics,

University of Science and Technology of China, Hefei, Anhui 230026, P.R. China

We present scanning tunneling microscopy studies of the LaOFeAs parent compound of iron pnictide superconductors. Topographic imaging reveals two types of atomically flat surfaces, corresponding to the exposed LaO layer and FeAs layer respectively. On one type of surface, we observe strong standing wave patterns induced by quasiparticle interference of two-dimensional surface states. The distribution of scattering wavevectors exhibits pronounced two-fold symmetry, consistent with the nematic electronic structure found in the $Ca(Fe_{1-x}Co_x)_2As_2$ parent state.

Similar to the high $T_C$ cuprates, the recently discovered iron pnictides also have layered crystal structures and complex phase diagrams [1]. Especially, the parent compounds of the pnictides undergo tetragonal to orthorhombic structural phase transition, accompanied by the formation of stripe-like antiferromagnetic (AF) state [2]. Both phase transitions reduce the four-fold ($C_4$) symmetry at high temperatures to two-fold ($C_2$) symmetry in the ground state. Superconductivity emerges as both orderings are weakened by doped charge carriers. Understanding the electronic structure of the ordered parent compound is thus an important task in unveiling the origin of the high $T_C$ superconductivity.

Scanning tunneling microscopy (STM) on the pnictides has provided atomic scale information regarding the surface morphology and electronic properties [3-8], vortex core states [9, 10], and superconducting pairing symmetry [11]. Most of these STM studies focused on the 122 pnictides $A$Fe$_2$As$_2$ ($A$ = Ca, Sr, or Ba) because single crystals with wide doping levels are available. A remarkable finding made by STM is the nematic electronic structure in the Ca(Fe$_{1-x}$Co$_x$)$_2$As$_2$ 122 parent state [12]. The existence of static electronic ordering along the orthorhombic $a$-axis and dispersive quasiparticles (qps) along the $b$-axis cannot be simply explained by the minute lattice distortion. A more complex theoretical model than previously expected is needed to describe the electronic structure of the pnictides.

However, analysis of the STM data on 122 pnictides is plagued by the $A$ atoms left on the FeAs layer upon cleaving [3-7, 10, 12]. The absence of this complication in the 1111 pnictides $Re$OFeAs ($Re$ = rare earth elements) make them attractive candidates for STM studies. More importantly, the 1111 compounds have the highest $T_C$ ~ 55K among all pnictides discovered so far [13]. In this letter we report STM studies of cleaved LaOFeAs parent compound. On one type of exposed surface, we observe strong standing wave patterns induced by quasiparticle interference (QPI) of two-dimensional (2D) surface electronic states. Fourier analysis shows that the **q**-space structure of QPI exhibits pronounced $C_2$ symmetry, reminiscent of the nematic electronic structure found in the 122 parent state [12].

High quality LaOFeAs single crystals are grown with NaAs flux method as described elsewhere [14]. Resistivity measurement shows a transition into the AF state at $T_N \sim 140K$. For STM experiments, LaOFeAs single crystals are cleaved at 77K in ultrahigh vacuum and then immediately transferred into the STM stage, which stays at 5K with pressure lower than $1\times10^{-10}$ mbar. All STM data reported here are acquired at 5.0 K.

Fig. 1 shows a large scale (800Å×800Å) topographic image of cleaved LaOFeAs, which reveals two types of clean and flat surfaces. The higher terrace in this image has a well resolved lattice structure and small cigar-shaped defects. The lower terrace, on the other hand, has ripple-like patterns around impurity centers and along the step edges. The two unequivalent surfaces in LaOFeAs are expected from its crystal structure, as schematically illustrated in Fig 1 inset. LaOFeAs consists of alternatively stacked LaO and FeAs layers with strong intralayer bonding and weak interlayer bonding. The crystal cleaves more easily between the LaO and FeAs layers (gray planes in Fig. 1 inset), resulting in the exposure of As-terminated FeAs layer and La-terminated LaO layer with equal possibilities. It is very difficult to distinguish the LaO layer and FeAs layer because the two layers have the same lattice structure and similar height [2]. For convenience in the following they are referred to as the "A" surface and "B" surface respectively, as indicated in Fig. 1.

Fig. 2a is a 200Å×200Å topographic image of the "A" surface, which has a clear atomic lattice decorated with cigar-shaped defects of unknown origin. The high-resolution image of a defect free area (Fig. 2a inset) shows that the lattice has a nearly perfect square symmetry with lattice constant around 4.0Å. We note that currently our STM cannot resolve the ~1% lattice distortion because it is beyond the accuracy of the piezo-element calibration. The 4.0Å lattice constant is in good agreement with the distance between neighboring La or As atoms in the original lattice [2]. Fig. 2b shows the *dI/dV* (differential conductance) spectrum of the "A" surface taken on a defect free region. The curve peaks sharply at positive sample bias $V = +100$mV, indicating a large electron density of state (DOS) in the unoccupied state.

Fig. 2c is a 200Å×200Å topographic image of the "B" surface. The most striking feature here is the existence of strong standing waves around impurity centers. The amplitude of the standing waves decays rapidly from the impurities. At relatively clean areas between impurities, the atomic lattice can still be resolved, as shown in Fig. 2c inset. The lattice also has a nearly square symmetry with lattice constant ~4.0Å, same as that in the "A" surface. The absence of structural reconstruction on both surfaces significantly simplifies the analysis of the STM data. Fig. 2d shows the *dI/dV* spectrum of the "B" surface taken at locations away from impurities. In contrast to that of the "A" surface, the spectrum here has a peak at negative sample bias $V = -200$mV, implying a large electron DOS in the occupied state.

The standing waves in the "B" surface are induced by the scattering interference of qps, as has been observed in various metal [15], semiconductor [16], and superconductor surfaces [17]. The interference between incident electron wave with wavevector $\mathbf{k}_i$ and elastically scattered electron wave $\mathbf{k}_f$ will lead to a spatial DOS oscillation with wavevector $\mathbf{q} = \mathbf{k}_f - \mathbf{k}_i$ and wavelength $\lambda = 2\pi/q$. Since the scattering wavevector $\mathbf{q}$ contains important $\mathbf{k}$-space information, the QPI imaging technique has become a powerful tool for studying the Fermi surface geometry and dispersion relation of novel materials [18, 19].

To extract the $\mathbf{k}$-space information from QPI, we take a series of differential conductance maps $dI/dV(\mathbf{r}, V)$ at different bias voltages, as shown in the upper panels of Fig. 3. The standing waves in each map correspond to the QPI of electrons with selected energy $E = eV$. Both the wavelength and spatial pattern of QPI evolve with $E$, manifesting the dispersion of qps. The lower panels of Fig. 3 show the Fourier transformed (FT) conductance maps $dI/dV(\mathbf{q}, V)$, which illustrate the $\mathbf{q}$-space structure of QPI. The nearly square atomic lattice renders four sharp Bragg peaks (surrounded by gray circles) with fixed wavevector $q_0 = 2\pi/4\text{Å}$. The scattering wavevectors responsible for the QPI, on the other hand, are the broad $\mathbf{q}$-space features that disperse with energy.

The most striking feature revealed by the FT maps is the pronounced $C_2$ symmetry exhibited by the distribution of scattering wavevectors. The dominant $\mathbf{q}$ points are distributed along two vertical arcs in the present orientation, apparently breaking the $C_4$ symmetry. There are also slightly weaker $\mathbf{q}$ points in the central strip along the vertical direction, but not along the horizontal direction. Since the $\mathbf{q}$-space feature derives from the $\mathbf{k}$-space structure, the $C_2$ symmetric $\mathbf{q}$ map strongly indicates that the $\mathbf{k}$-space electronic structure of the LaOFeAs surface has $C_2$ symmetry. To identify the orientation of the $\mathbf{q}$ vectors, the $\mathbf{q}$ map taken at $E = -125$meV is plotted in Fig. 4a with respect to the Brillouin zone (BZ) of the pnictides. The broken square indicates the unreconstructed BZ and the solid square indicates the reconstructed BZ of the parent state. By comparison to the Bragg peaks of the As (or La) lattice, we found that the arcs with dominant $\mathbf{q}$ points are aligned along the Fe-Fe bond direction, although we cannot distinguish the orthorhombic *a* and *b* crystal axis. The overall feature bears strong resemblance to the FT conductance maps of parent $Ca(Fe_{1-x}Co_x)_2As_2$ [12].

The QPI on LaOFeAs are most likely induced by 2D surface electronic states, rather than from the bulk electronic states as claimed in $Ca(Fe_{1-x}Co_x)_2As_2$ [12]. The standing wave patterns are so pronounced that they dominate the topographic image. The atomic lattices are almost concealed by the standing waves on the "B" surface. Similar behavior has only been found in materials with well defined surface electronic structures [15, 16]. The QPI of the bulk states, conversely, shows much weaker spatial modulations that can only be visualized when the lattice effect is removed [12]. Surface states on cleaved 1111 pnictides were first hinted by angle resolved photoemission spectroscopy (ARPES), which found extra large hole pocket around the Γ point incompatible with bulk band structure [20-22]. Density functional theory (DFT) calculations also suggest the formation of 2D surface states on the highly polar surfaces of cleaved LaOFeAs [23]. Recent ARPES data provide direct evidence for the existence of surface states by showing the lack of $k_z$ dispersion for certain bands on cleaved LaOFeAs [24, 25].

The spatial resolution of STM reveals new features about the surface states. Both DFT calculation [23] and ARPES measurements [25] suggest that surface states exist on both LaO and FeAs layers. However, in our STM results surface states are only found in one type of surface and are totally absent in the other. This observation suggests that the surface electronic structure is strongly affected by the underneath bulk states, although the exact difference between the two situations needs to be further investigated. In fact, ARPES measurements on LaOFeAs find that the surface state qp scattering rate is significantly reduced as the bulk enters the AF state [25], implying sizable coupling between the surface and bulk states.

The $C_2$ symmetry of the surface state is another finding owing to the local ability of STM, which allows us to observe a single orthorhombic domain. The similarity between the **q**-space structure of LaOFeAs and the nematic electronic structure in $Ca(Fe_{1-x}Co_x)_2As_2$ suggests that nematicity is a rather universal feature in the parent state of the pnictides. As has been discussed extensively, nematicity is a central issue of the pnictide electronic structure, and may hold the key to understanding the mechanism of superconductivity [26-29]. However, the $C_2$-symmetric surface states reported here display an important difference from the nematicity in 122 [12]. In $Ca(Fe_{1-x}Co_x)_2As_2$, the electronic structure is static along the orthorhombic *a*-axis and dispersive along the *b*-axis [12], which is more consistent with the melted stripe origin of the nematicity [30]. In the surface states of LaOFeAs, however, the behavior is quite different. Fig 4b shows the *E* vs. *q* relations for **q** along the two orthorhombic Fe-Fe bond directions, as indicated by the arrows in Fig. 4a. Apparently the qps are dispersive along both directions, which is more consistent with the Pomeranchuk instability picture of the nematic electronic structure [28], rather than the melted smectic order scenario. Of course we cannot exclude the subtle differences between the surface and bulk states, and between the 1111 and 122 systems.

Another surprising fact is the similarity between the dispersions along the $q_1$ and $q_2$ directions shown in Fig. 4b. We would expect a strong anisotropy between them given the stripe-like AF order at this temperature and the apparent $C_2$ symmetry of the overall $q$-space structure. However, we also note that the dispersion relationships are quite complicated. They are close to linear at low energies and become less dispersive at high energies, thus cannot be fit by a simple parabolic relation $E \sim q^2$ like that in Cu(111) [15]. The deviation from a single electron band dispersion is probably due to scatterings between different Fermi surface pockets, which have been observed on the 1111 surface by ARPES [23, 25]. In a multiband situation it is in general very difficult to deduce the $k$-space structure from the $q$ map alone. Further theoretical investigations combining the ARPES and STM results may help to elucidate these puzzles.

In summary, STM studies on LaOFeAs parent compound uncover QPI patterns caused by 2D surface states. The $C_2$ symmetry of the surface electronic structure is consistent with the nematic electronic order found in the 122 parent state, although discrepancies do exist between the two systems. These results are valuable for interpreting experimental data obtained by surface sensitive probes on 1111 pnictides. They also shed new light on the possible mechanism of nematicity in the parent states of pnictides.

We thank D.L. Feng, J.P. Hu, D.H. Lu, Z.Y. Weng, C.K. Xu, H. Yao, H. Zhai, and X.J. Zhou for helpful discussions. This work was supported by the National Natural Science Foundation of China (Grant 10834003, 10925417) and by the Ministry of Science and Technology of China (Grant 2006CB601001, 2006CB922005, and 2009CB929402).

* *These authors contributed equally to this work.*

† Electronic address: yayuwang@tsinghua.edu.cn

**Figure Captions:**

FIG 1. 800Å×800Å constant current image of cleaved surface of parent LaOFeAs taken with sample bias $V = -100$mV and tunneling current $I = 20$pA. The two different types of surfaces are labeled as "A" and "B" respectively. (Inset) Schematic crystal structure of LaOFeAs.

FIG 2. (a) 200Å×200Å topography of "A" surface taken with $V = -200$mV and $I = 50$pA. (Inset) 40Å×40Å close-up image shows clear atomic lattice with 4Å lattice constant. (b) $dI/dV$ of the "A" surface has a peak at positive sample bias around $+100$mV. (c) 200Å×200Å topography of "B" surface taken with $V = -100$mV and $I = 40$pA displays strong standing waves around scattering centers. (Inset) 40Å×40Å image shows the atomic lattice. (d) $dI/dV$ of the "B" surface has a peak at negative sample bias around $-200$mV.

FIG 3. (upper panel) Differential conductance maps $dI/dV(\mathbf{r}, V)$ over a 400Å×400Å area on "B" surface taken with sample bias $-125$mV, $-75$mV, $75$mV, and $125$mV. (lower panel) Fourier transformed conductance maps $dI/dV(\mathbf{q}, V)$ reveal that the $\mathbf{q}$-space structure of QPI exhibits pronounced $C_2$ symmetry.

FIG 4. (a) The $\mathbf{q}$ map taken at $-125$mV with respect to the Brillouin zone of LaOFeAs. The broken (solid) square indicates the unreconstructed (reconstructed) BZ. The dominant QPI wave vectors are distributed along the Fe-Fe bond direction. The Bragg peaks with $q_0 = 2\pi/a_{\text{As-As}} = 2\pi/4$Å are circled in gray. (b) The $E$ vs. $q$ dispersions for the two $\mathbf{q}$ wavevectors along the Fe-Fe bond directions as indicated by arrows in (a).

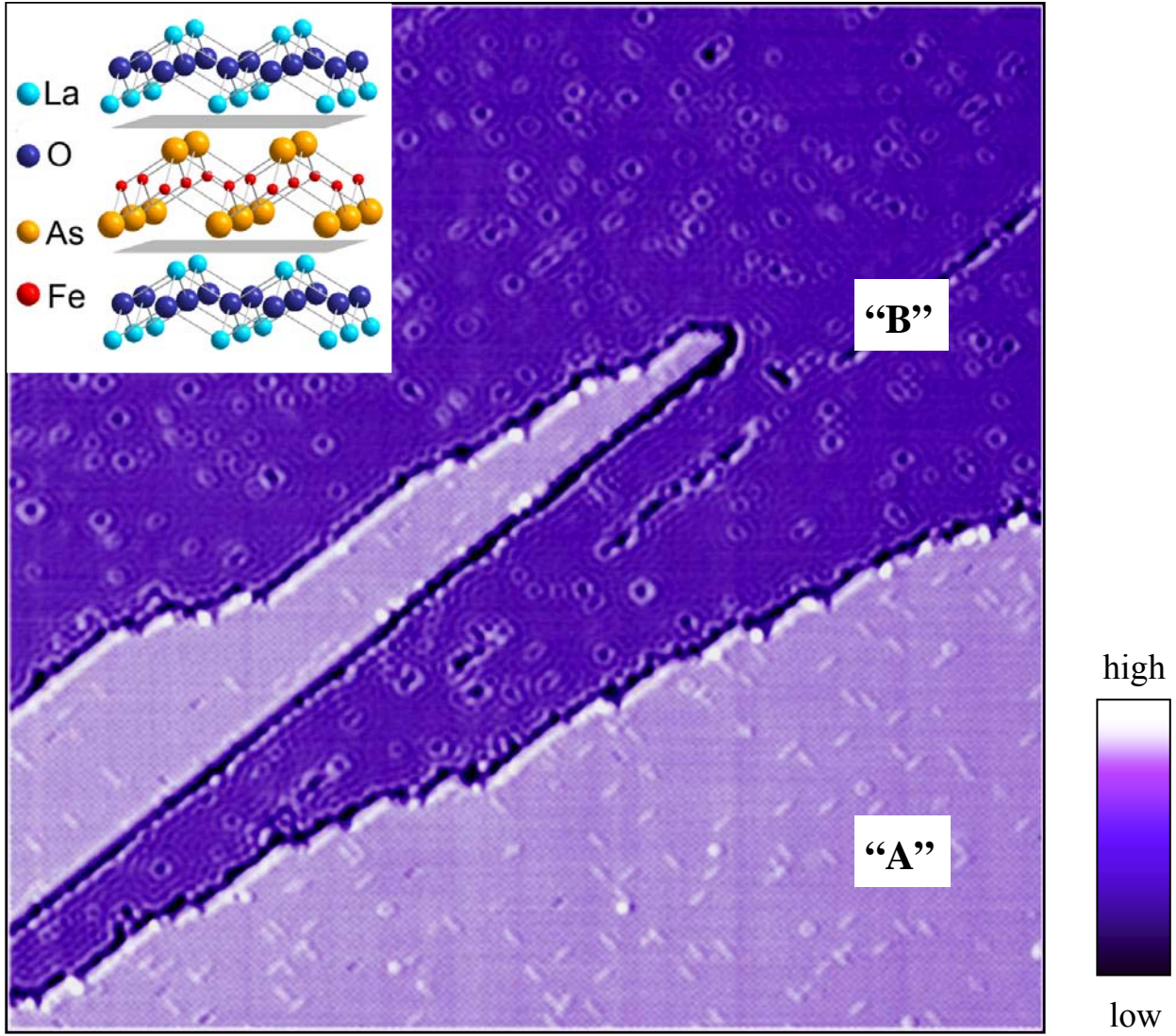

Figure 1

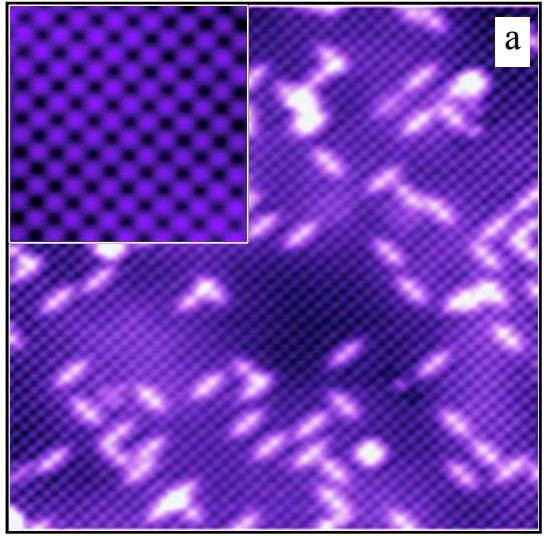
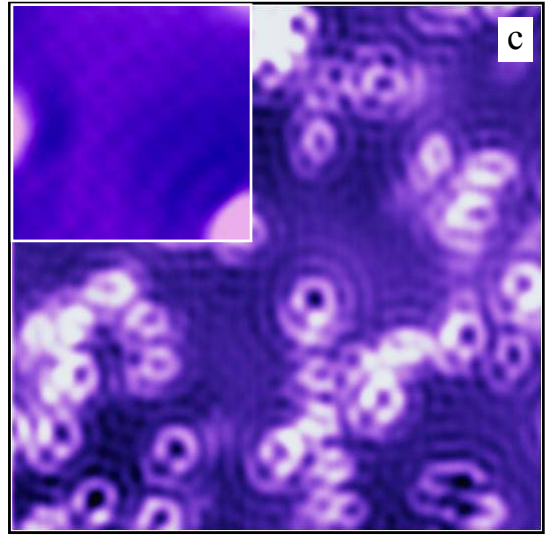
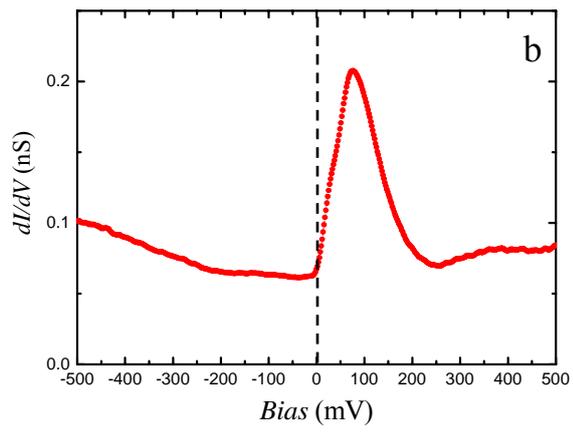
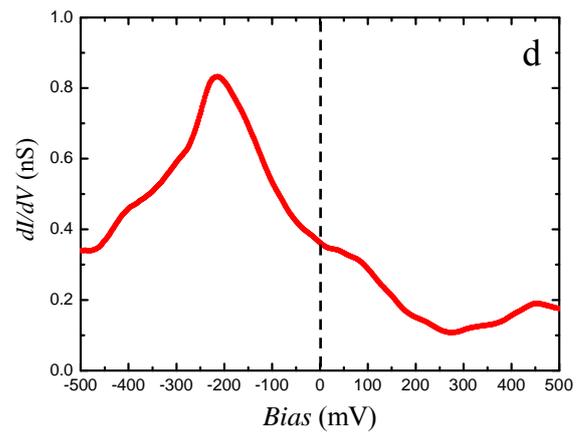

Figure 2

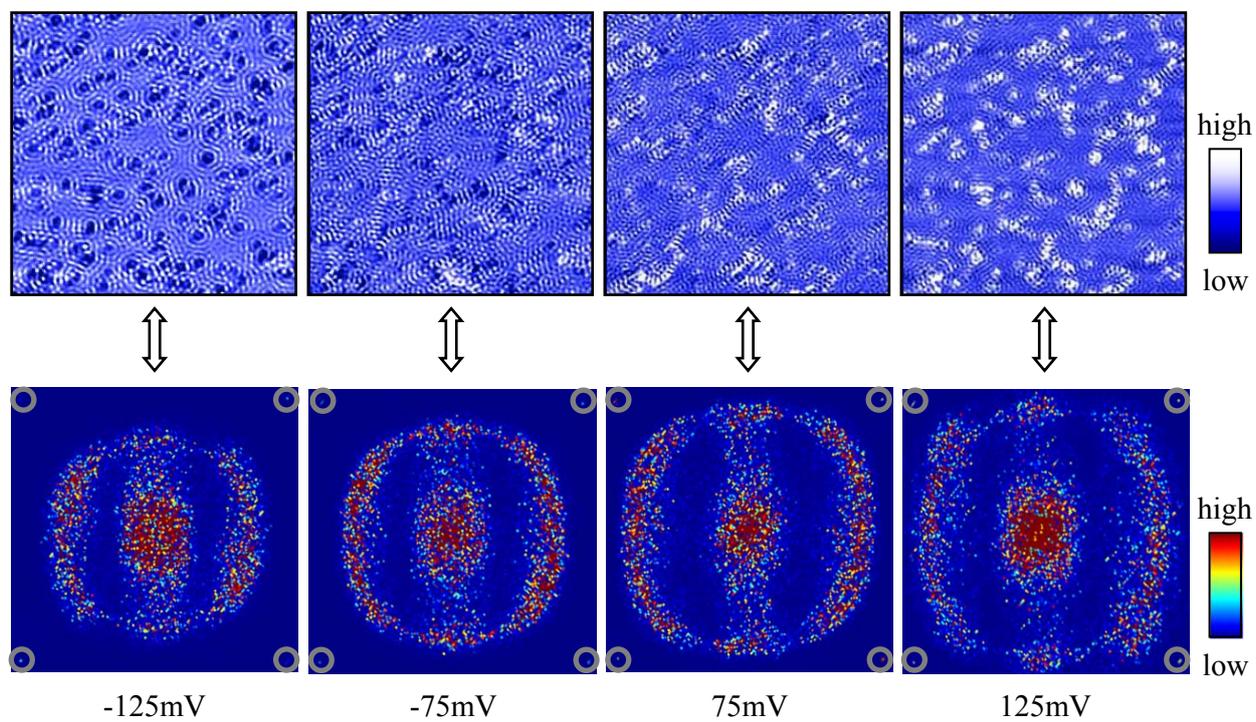

Figure 3

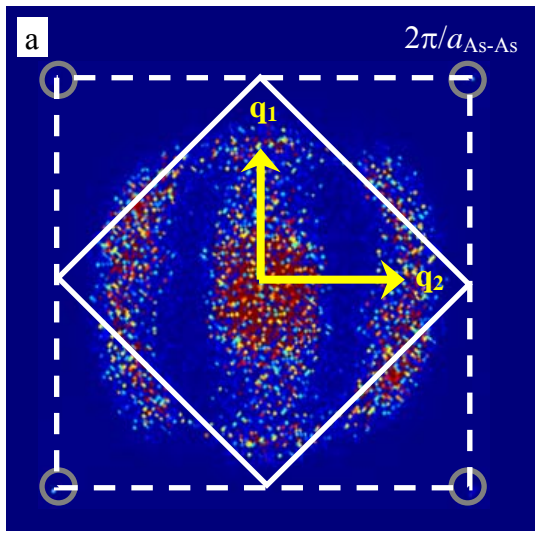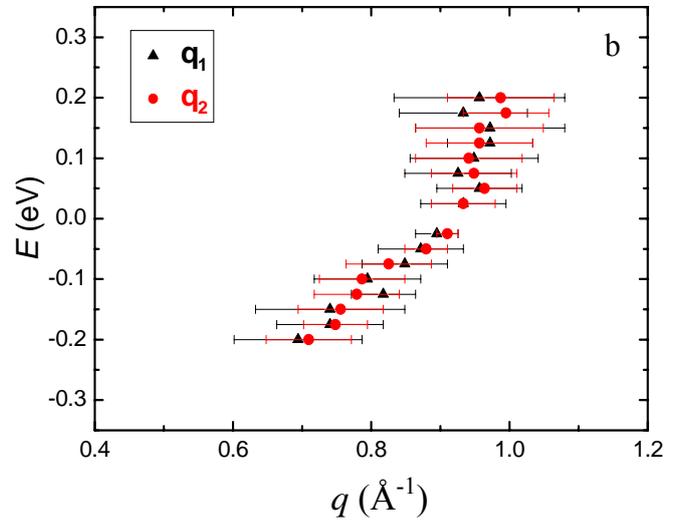

Figure 4